\documentclass[a4paper,fleqn,usenatbib]{mnras}

\usepackage{txfonts}
\usepackage[T1]{fontenc}
\usepackage{ae,aecompl}

\usepackage{graphicx}	
\usepackage{amssymb}	

\title[Study of the filament]{Formation and material supply of an active-region filament associated with newly emerging flux}

\author[Wang et al.]{
Jincheng Wang,$^{1}$\thanks{E-mail: wangjincheng@ynao.ac.cn}
Xiaoli Yan,$^{1,2}$
Qiaoling Guo,$^{3}$
Defang Kong,$^{1,2}$
Zhike Xue,$^{1,2}$
\newauthor
Liheng Yang,$^{1,2}$
and Qiaoling Li$^{1,4}$
\\
$^{1}$Yunnan Observatories, Chinese Academy of Sciences, Kunming 650011, People.s Republic of China.\\
$^{2}$Center for Astronomical Mega-Science, Chinese Academy of Sciences, 20A Datun Road, Chaoyang District, Beijing, 100012,\\
People.s Republic of China.\\
$^{3}$College of Mathematics Physics and Information Engineering, Jiaxing University, Jiaxing 314001, People's Republic of China. \\
$^{4}$University of Chinese Academy of Sciences, Yuquan Road, Shijingshan Block Beijing 100049, People.s Republic of China.
}

\date{Accepted 2019 July 08. Received 2019 June 29; in original form 2019 April 18}

\pubyear{2019}

\begin{document}
\label{firstpage}
\pagerange{\pageref{firstpage}--\pageref{lastpage}}
\maketitle

\begin{abstract}
With the observations of $SDO$/AIA 304 $\rm\AA$ and NVST H$\alpha$ bands, we present the formation process of an active-region filament in active region NOAA 11903 during the period from 02:00 UT to 10:00 UT on November 25, 2013. A series of jets occurring in the vicinity of the south-western footpoint of the filament directly ejected cool and hot plasma to filament height and supplied material for the filament. Some newly emerging flux is found in the vicinity of the filament south-western footpoint during these jets. In this paper, we mainly focus on the material supply of the filament. The plasma mass uploaded by the jets and the mass of the filament are estimated, which manifests the fact that the mass carried by jets can supply sufficient material for the formation of the filament. We found two types of jets. one is H$\alpha$ jet, and the other is EUV jet. The significant finding is that some cool jets seen in the H$\alpha$ band but not in the SDO/AIA bands also could eject the cool material for the filament. These results suggest that cool plasma in the low atmosphere can be directly injected into the upper atmosphere and become the filament material by two types of jets. Moreover, the newly emerging flux with non-potential field plays an important role in the appearances of the jets and the magnetic structure of the filament.
\end{abstract}

\begin{keywords}
Sun:filaments - Sun:atmopshere - Sun:activity - emerging flux
\end{keywords}



\section{Introduction} \label{sec:intro}
Intriguing structures in the solar atmosphere, solar filaments are sheets of relatively cool and dense material structures suspended in the surrounding hot corona. They are bright cloud-like structures when observed beyond the solar limb called prominences and appear as dark filamentary objects when seen against the solar disk called filaments. Filaments are primarily found in the solar quiet region (quiescent filament), between active regions or surrounding them (intermediate filaments) and inside active regions (active-region filament) \citep{mac10}. Generally, active-region filaments are shorter, smaller, lower in height, and have shorter lifetimes than quiescent and intermediate filaments. Nevertheless, active-region filaments are more unstable and more likely to erupt \citep{par14}. As the filaments lose the equilibrium, they would erupt and eject the material into the high corona. Filament eruptions are often accompanied with other solar activities (such as solar flares, coronal mass ejections (CMEs), coronal jets, surges, and so on) \citep{jin04}.

Recently, understanding of solar filaments is one of the hot topics in solar physics, which includes their distribution, magnetic structure, formation, and eruption. Filaments always lie above the magnetic polarity inversion lines (PILs) separating opposite polarities of the photospheric magnetic field, which are considered to support by the local magnetic fields (e.g. magnetic dips or twisted structures) against the gravity \citep{bab55,mar98,che14a}. Unfortunately, due to the limitation of magnetic field measurements at present, it is difficult to measure directly the local magnetic fields in the filament. Therefore, the magnetic structures of filaments are still under controversy. There are two popular views on the magnetic topologies of filaments. One view is the shear-arched model \citep{kip57,ama91,ant94,wel05} and the other is the flux rope model \citep{kup74,dem89,yang14,yan15,che17}. Using two-dimensional magnetohydrodynamic (MHD) numerical simulations, \cite{che00} proposed reconnection-favored emerging flux could cancel with the magnetic field below the flux rope or the magnetic field on the outer edge of the channel and led to the eruption of the flux rope. Torus instability and Kink instability are also proposed to be the triggers for the eruption of the filaments \citep{tor03,kli06,jia13,yan14,li16}.

Formation of the filaments, which consists of the formation of magnetic structures and the mass supply, is still controversial. Regarding of the first term (formation of the filament magnetic structures), many researchers have suggested two different mechanisms to form magnetic structures of filaments: surface effect \citep{van89,mar01,yan15,yan16,wan17,xue17,che18} and subsurface effect \citep{rus94,oka08,oka09,mact10,lit09,lit10,yan17}. In the former, the magnetic structures of filaments are formed gradually through the magnetic reconnection driven by various photospheric motions (such as sheared flow, converging flow, and rotating motion). In the latter, the magnetic structures of filaments are thought to be formed in the solar interior and emerge into the atmosphere through magnetic buoyancy. Although both views can explain some observational characteristics, the complicated processes of magnetic structures of filament formation are still not fully understood.

Mechanism of the filament mass supply is another issue for understanding the formation of the filament. It is believed that the materials of filaments originate from low atmosphere (such as from solar chromosphere) \citep{pik71,zir94,son17}, which are transported to solar corona either through magnetic forces or thermal pressure forces. Based on observational characteristics and numerical simulations, three popular models have been proposed: injection model \citep{wan99,liu05,cha03,wan18}, levitation model \citep{rus94,gal99,kuc12}, and evaporation-condensation model \citep{ant91,kar05,xia11,xia12,liu12}. In the injection model, cool plasma in the low atmosphere is forced upward in the filament through sufficient force in the magnetic reconnection. Using the UV and EUV data from \emph{Transition Region and Coronal Explorer (TRACE}), \cite{cha03} reported that a series of jets and two eruptions events could supply the mass necessary for the formation of a reverse S-shaped filament in active region NOAA 8668. \cite{zou16} also found that cool material could be injected into the filament spine with a speed of 5-10 km/s and magnetic reconnection played an important role in these transport processes. In the levitation model, cool plasma is lifted by rising magnetic fields at the PIL, which resides in the upward concavities of the helical field (magnetic dips structure). In the evaporation-condensation model, chromospheric plasma is evaporated and flow up driven by heating localized nearby the footpoint, which ultimately condenses in the corona as cool prominence mass. Due to the limitation of the observation, it is still controversial on these three models. In other words, poor observational evidence makes it difficult for us to understand fully the mechanism of filament mass supply.

In order to understand the formation of the filament, we study the formation process of an active region filament in NOAA 11903 observed by \emph{Solar Dynamics Observatory} and New Vacuum Solar Telescope during the period from 02:00 to 10:00 on November 25, 2013. We mainly investigate the material injection process of the filament. The sections of this paper are organized as follows: the observations and methods are described in Section \ref{sec:obser methods}, the results are presented in the Section \ref{sec:results}, and the main conclusion and discussion are given in Section \ref{sec:conclusion}.
\section{Observations and Methods} \label{sec:obser methods}
\subsection{Observations}
The data set are primarily from New Vacuum Solar Telescope\footnote{\url{http://fso.ynao.ac.cn}} (NVST; \citealp{liu14}) and \emph{Solar Dynamics Observatory}\footnote{\url{https://sdo.gsfc.nasa.gov}} (\emph{SDO}; \citealp{pes12}). The NVST is a vacuum solar telescope with a 985 mm clear aperture, located at Fuxian Lake, in Yunnan Province, China. It is mainly composed of four instrumentation systems: the adaptive optics system, the polarization analyzer, the imaging system, and the spectrometers. The imaging system has three high-resolution imaging channels for monitoring the photosphere and chromosphere, which include two channels for photosphere in TiO (7058 {\AA}) and G-band (4300 {\AA}) and one channel for chromosphere in H\rm{$\alpha$} (6562.8 {\AA}) band. The high-resolution H\rm{$\alpha$} images are utilized to study the filament formation process in this paper, and their cadence and spatial resolution are 12 s and $0\farcs163$ per pixel, respectively. The Atmospheric Imaging Assembly \citep{lem12} and the Helioseismic and Magnetic Imager (HMI; \citealp{sche12}; \citealp{sch12}) on board the \emph{SDO} provide full-disk, multiwavelength, high spatio-temporal resolution observations for this study. The \emph{SDO}/AIA provides seven extreme ultraviolet and three ultraviolet-to-visible channel images with a spatial resolution of 0$\farcs$6 per pixel and a minimum cadence of 12 s, while the \emph{SDO}/HMI provides line-of-sight magnetic fields, continuum intensity, Doppler shift, and vector magnetograms (VMs) on the photosphere with a spatial resolution of 0$\farcs$5 per pixel, with the cadences of the three former channels being about 45 s and that of the latter one being about 12 minutes.
\subsection{Methods}
We calculate the negative magnetic flux and the change rate of magnetic flux as in the flowing equations:
\begin{equation}\label{eqflux}
    \phi = \int B_{zn} dA,
\end{equation}
and
\begin{equation}\label{rate}
    r = \frac{d|\phi|}{dt},
\end{equation}
in which $B_{zn}$ denotes the negative magnetic field, A is the integrative areas and t is the time.

According to the Ampere's law, the current density can be derived by follow equation:
\begin{equation}\label{eq1}
    \textbf{J} = \frac{1}{\mu_0}(\bigtriangledown \times \textbf{B}),
\end{equation}
in which $\mu_0$ is the magnetic permeability of vacuum and \textbf{B} is the magnetic field vector. Therefore, the vertical current density perpendicular to the solar surface, can be calculated by using the $SDO$/HMI VMs data. Based on the equation:
\begin{equation}\label{eq2}
    j_z = \frac{1}{\mu_0}(\bigtriangledown \times \textbf{B})_z = \frac{1}{\mu_0}(\frac{\partial B_y}{\partial x} - \frac{\partial B_x}{\partial y}),
\end{equation}
where $B_x$ and $B_y$ are the two perpendicular components of the transverse magnetic fields. With the data from $SDO$/HMI VMs, the distribution of $j_z$ on the solar surface can be obtained every 12 minutes. Therefore, the vertical current can be integrated by the vertical current density in the interesting regions.

Magnetic helicity and magnetic energy can be transported from the solar interior to the corona by the new emergence flux and the various motions of magnetic flux on the photosphere. The injection rate of the magnetic helicity across a surface S is expressed by \citep{ber84,dem03,par05}:
\begin{equation}\label{eq3}
   \dot H = 2\int_S(\textbf{A}_p\cdot \textbf{B}_t)V_{\bot n}dS - 2\int_s(\textbf{A}_p\cdot \textbf{V}_{\bot t})B_n dS,
\end{equation}
where $\textbf{A}_p$ is  the vector potential of the potential field $\textbf{B}_p$, $\textbf{B}_t$ and $\textbf{B}_n$ denote the tangential and normal magnetic fields, $\textbf{V}_{\bot t}$ and $V_{\bot n}$ are the tangential and normal components of the velocity perpendicular to the magnetic field lines, respectively. We derive the vector velocity field by using the Differential Affine Velocity Estimator for Vector Magnetograms (DAVE4VM) method \citep{sch08}. The first term contributed by the emerging flux tubes is named emergence term, and the second term generated by the shearing and braiding the field lines by the tangential motions is named shear term. Similarly, the rate of the magnetic energy (Poynting vector) can be derived by following equation \citep{kus02,bi18}:
 \begin{equation}\label{eq4}
    \dot E = \frac{1}{4\pi}\int _s B_t^2V_{\bot n}dS - \frac{1}{4\pi}\int_s(\textbf{B}_t \cdot \textbf{V}_{\bot t})B_ndS.
 \end{equation}
In the same way, the first term of the equation is named emergence term, which is associated with the emergence of twisted magnetic tubes from the solar interior. The second term of the equation is named shear term, which is generated by shearing magnetic field lines due to tangential motions on the surface. Accumulative helicity and energy can be derived by the integration of the helicity injection rate and energy injection rate with the time, respectively. To reduce the noise influence from the measurement, we only consider the field with a transverse component stronger than 150 G. We use these non-potential physical parameters to investigate the newly emerging flux related to the mass supply of the filament.
\section{Results} \label{sec:results}
\subsection{Formation process of the active-region filament}
\begin{figure}
\includegraphics[width=\columnwidth]{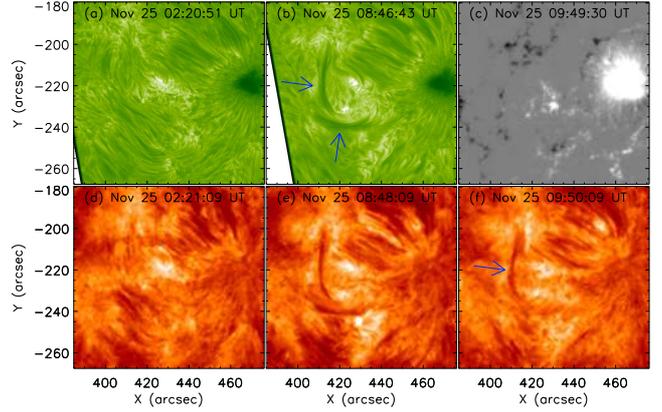}
\caption{Formation process of the filament. (a)-(b) H$\alpha$ images observed by NVST at different moments. (c) The line-of-sight magnetic fields from $SDO$/HMI. White patches denote the magnetic field with positive polarity, while black ones denote the magnetic field with negative polarity. (d)-(f) The corresponding 304 $\rm\AA$ images observed by $SDO$/AIA. \label{fig1}}
\end{figure}

 A filament formed gradually in active region NOAA 11903 during the period from 02:00 UT to 10:00 UT. Active region NOAA 11903 located at the southwestern hemisphere (e.g., (460$\arcsec$, -220$\arcsec$)) comprises a positive dominant sunspot and some discrete negative magnetic fields on November 25, 2013. The formation position of the filament was close to the dominant sunspot (see Fig.\ref{fig1}). The panels (a) and (b) of Fig.\ref{fig1} show the formation process in H$\alpha$ observations from NVST. At 02:20:51 UT, the filament was absent in the field of view. At 08:46:43 UT, two filamentary structures had formed, which were marked by two blue arrows. Panel (c) of Fig.\ref{fig1} shows the line-of-sight (LOS) magnetic field from $SDO$/HMI at 09:49:30 UT, which manifests that the south-western footpiont of the filament roots in the positive magnetic field and the north-eastern footpoint roots in the negative magnetic field. Panels (d)-(f) of the Fig.\ref{fig1} exhibit the formation process of the filament in 304 $\rm\AA$ images observed by $SDO$/AIA. At 09:50:09 UT, a filament could be seen in the place marked by the blue arrow in the panel (f) of Fig.\ref{fig1}. The detailed evolutionary formation process of this filament is shown by the animation of Fig.\ref{fig1}. According to the animation, we find that a series of jets driven the cool plasma or chromospheric material to filament height. Unidirectional material motions driven by the jets besides the south-western footpoint of the filament were found and then the filament generally appeared. It should be noted that some of this plasma could remain at the filament height and contribute to the mass of the filament, while some of this plasma could instead fall down to the chromosphere.

\begin{figure}
\includegraphics[width=\columnwidth]{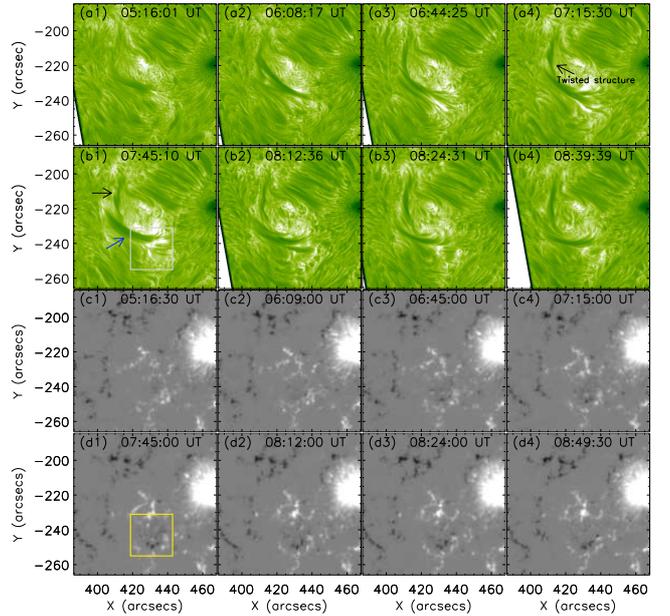}
\caption{The material injection process of the filament and the evolution of photospheric magnetic fields. (a1)-(a4) and (b1)-(b2) H$\alpha$ images at different moments observed by NVST. The black arrow and the blue one indicate two filamentary structures in the panel (b1), respectively. (c1)-(c4) and (d1)-(d4) The corresponding line-of-sight field from $SDO$/HMI. The yellow box in the panel (d1) and the gray box in the panel (b1) show the same region for calculating negative flux and intensities of $SDO$/AIA seven bands.\label{fig2}}
\end{figure}

Fig.\ref{fig2} shows numerous cool plasma ejected into the filament height by a series of jets. Panels (a1)-(a4) and (b1)-(b4) exhibit formation process of the filament in H$\alpha$ images. A series of jets occurred in the vicinity of the south-western footpoint of the filament. Accompanying the appearance of these jets, the cool plasma was lifted into the filament height and some of them became the filament material. Two filamentary structures could be identified in panels (b1)-(b4), which are marked by the black arrow and the blue one. The plasma injected by the jets went along the magnetic field lines in the filamentary structure marked by the blue arrow and then fall down to anther footpoint, while the plasma could remain in the filament structure marked by the black arrow. Eventually, the filamentary structure marked by the black arrow became the maturing filament (see the panel (f) of the Fig.\ref{fig1}). An intriguing feature is that the twisted structure can be distinguished in one filamentary structure marked by the black arrow (see panels (a4), (b1)-(b3)). The twisted structure is marked by the back arrow in the panel (a4). However, the other filamentary structure does not exhibit any twisted structures. This means twisted magnetic fields exist in the filamentary structure marked by the black arrow and the other filamentary structure lacks of the twisted magnetic field. This also can explain why the lifted plasma could not remain in the filamentary structure marked by the blue arrow, which is due to the lack of twisted structure or dips structure in that located magnetic field. Panels (c1)-(c4) and (d1)-(d4) show the corresponding images of the line-of-sight magnetic fields. White patches denote the magnetic field with positive polarity, while black ones denote the magnetic field with negative polarity. The adjacent main sunspot was the positive magnetic field. Based on these panels, one can see that the emergence of some magnetic fluxes occurred in the vicinity of the filament marked by the yellow box in the panel (d1) of Fig.\ref{fig2}. Therefore, we suspect that there is a close relationship between the jets and the flux emergence. More and more observational evidences and numerical simulations manifest that the jets can be triggered by the magnetic reconnection between the emerging flux and the pre-existing magnetic flux \citep{yok95,shi98,cheu14,li17}.
\subsection{The jets and material supply for the filament} \label{sec3.2}
\begin{figure}
\includegraphics[width=\columnwidth]{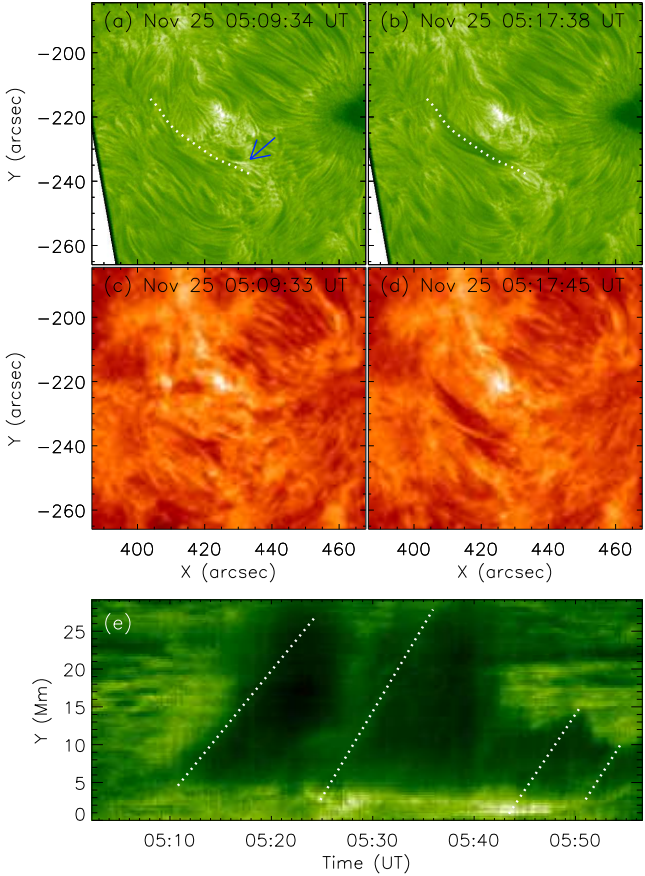}
\caption{The material injection event observed by the NVST. (a)-(b) H$\alpha$ images observed by NVST at 05:09:34 UT and 05:17:38 UT. The blue arrow in the panel (a) indicates the brightening in H$\alpha$, while the white dotted lines indicate the same path for the time-distance diagrams in pane (e). (c)-(d) $SDO$/AIA 304 $\rm\AA$ images at corresponding time. (e) The time-distance diagrams derived from a series of H$\alpha$ images from 05:02 UT to 05:56 UT. \label{fig3}}
\end{figure}

The cool material began to be lifted by a small jet (H$\alpha$ jet) to filament height at about 05:09 UT. Fig.\ref{fig3} (a) and (b) show this material injection event in H$\alpha$ images observed by the NVST. At 05:09:34 UT, we find a brightening occurred in the vicinity of the south-western footpoint of the filament marked by the blue arrow in the panel (a), and the material began to be injected from the chromosphere to higher atmosphere. The panel (b) shows that the cool plasma was directly injected into the filament and the intensity in the vicinity of the south-western footpoint had a slight enhancement. Fig.\ref{fig3} (c) and (d) show the corresponding 304 $\rm\AA$ images observed by $SDO$/AIA instrument. It is difficult to distinguish any brightenings in the corresponding place where the brightening was found in H$\alpha$ band. This mean that this jet is H$\alpha$ jet instead of EUV jet. It is most likely to be caused by the different responding temperature between AIA 304 $\rm\AA$ band images and H$\alpha$ band images in the region with abundant plasma. In order to understand the property of the injected cool plasma, we make a time-distance diagram to estimate the velocity of the ejected cool plasma along the white dotted lines in the panels (a) and (b). Panel (e) shows the time-distance diagram derived by a series of the images from the NVST H$\alpha$. From this time-distance diagram, we distinguish four inclined lines. Based on their inclinations, we derive the projection velocities along the four dotted white lines, which are about 27.4 km/s, 37.8 km/s, 33.4 km/s, and 35.1 km/s, respectively. The mean projection velocity is about 33.4 km/s.

\begin{figure}
\includegraphics[width=\columnwidth]{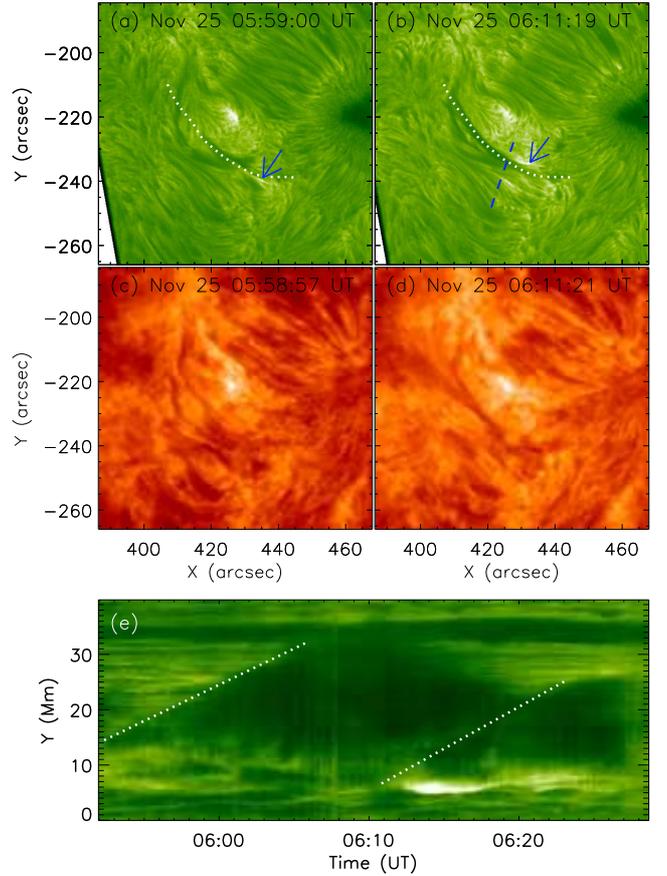}
\caption{The material injection event observed by the NVST. (a)-(b) H$\alpha$ images observed by NVST at 05:59:00 UT and 06:11:19 UT. The blue arrows in the panel (a) \& (b) indicate the brightenings in H$\alpha$, while the white dotted lines indicate the same path for the time-distance diagram in pane (e). The blue dashed line in panel (b) denotes the position of the time-distance diagram in panel (c) of Fig.\ref{fig5}. (c)-(d) $SDO$/AIA 304 $\rm\AA$ images at corresponding time. (e) The time-distance diagrams derived from a series of H$\alpha$ images from 05:51 UT to 06:28 UT. \label{fig4}}
\end{figure}

Fig.\ref{fig4} exhibits another material injection event during the period from about 05:51 UT to 06:35 UT. Panels (a) and (b) show this material injection event in H$\alpha$ images observed by the NVST at 05:59:00 UT and 06:11:19 UT, respectively. We also find some brightenings in the vicinity of the filament south-western footpoint, which are marked by the blue arrows. However, we also can not distinguish any brightenings at the corresponding places in $SDO$/AIA 304 $\rm\AA$ images in panels (c) and (d). This jet is also associated with H$\alpha$ jet instead of EUV jet. This means the cool plasma were ejected by the jet and injected into the filament, which is similar to the previous event. We also make a time-distance diagram to estimate the velocity of the ejected cool plasma along the white dotted lines in the panels (a) and (b). Panel (e) shows the time-distance diagram derived by a series of the images observed by the NVST H$\alpha$ band from 05:51 UT to 06:35 UT. According to this time-distance diagram, the projection velocities along the two dotted white lines are about 22.1 km/s and 24.1 km/s, respectively. The mean projection velocity is about 23.1 km/s.

\begin{figure}
\includegraphics[width=\columnwidth]{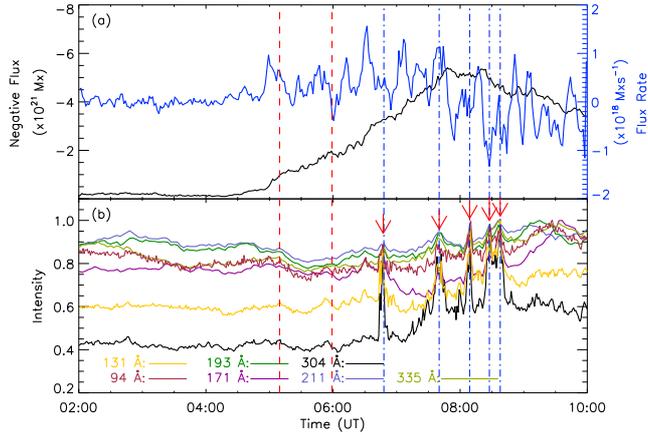}
\caption{Time profiles of different parameters in the yellow box in the panel (d1) of Fig.\ref{fig2}. (a) Evolutions of negative flux and flux rate. (b) The variations of $SDO$/AIA intensities in different bands. The red vertical dashed lines denote the onset of the first material injection event and the other material injection observed by the NVST at 05:09 UT and 05:59 UT, respectively. The blue vertical dashed-dotted lines indicate the time of EUV jets \label{fig5}}
\end{figure}

In order to investigate these jets related to the material injection of the filament, we calculate the intensities of the $SDO$/AIA seven bands (94 $\rm\AA$, 131 $\rm\AA$, 171 $\rm\AA$, 193 $\rm\AA$, 211 $\rm\AA$, 304 $\rm\AA$, 335 $\rm\AA$) integrated in the yellow box in the panel (d1) of Fig.\ref{fig2}. In addition, we also calculate the negative flux in the yellow box in the panel (d1) of Fig.\ref{fig2}, which is due to the complexity of the positive magnetic field nearby the south-western footpoint of the filament. Fig.\ref{fig5} (a) shows the time variations of negative flux and its increasing rate ($d|\phi|/dt$) during the period from 02:00 UT to 10:00 UT on November 25, 2013. The first vertical red dashed line indicates the onset of the material injection derived by the NVST, which is about at 05:09 UT on November 25. The negative flux enhanced gradually about from - 0.2 $\times$ $10^{21}$ Mx to - 5.5 $\times$ $10^{21}$ Mx during the period about from 05:00 UT to 07:45 UT, and the mean rate of enhancement is about 4.3 $\times$ $10^{17}$ Mx $\rm s^{-1}$. This means the new flux emerged from the subsurface in the vicinity of the south-western footpoint of the filament during the period from 05:00 UT to 07:45 UT. During the period from 07:45 UT to 10:00 UT, the negative flux decreased gradually from - 5.5 $\times$ $10^{21}$ Mx to - 3.5 $\times$ $10^{21}$ Mx, and the mean rate of decrease is about 2.4 $\times$ $10^{17}$ Mx $\rm s^{-1}$. The front four jets occurred in the phase of the flux enhancement, while the latter three jets occurred in the phase of flux decrease. On the other hand, two H$\alpha$ jets occurred in the lower magnitude of the magnetic flux, while the EUV jets occurred in the higher magnitude of the magnetic flux. This manifests that the H$\alpha$ jets were caused by the weak emerging flux and the EUV jets were caused by the strong emerging flux and the type of the jets is associated with the strength of the emerging flux. The blue line in the panel (a) indicates the increasing rate of the negative flux as the function of the time. One H$\alpha$ jet and two EUV jets had positive flux rate. Another H$\alpha$ jet and a EUV jet had negative flux rate. Two jets almost occurred at the around zero increasing rate. Fig.\ref{fig5} (b) shows the time profiles of normalized intensity in different bands of $SDO$/AIA. The normalized intensity is defined as that the maximum of intensity divided by intensity in each band. We can unambiguously distinguish several intensity enhancements during the period of the flux emergence, which are more sensitive in 304 $\rm\AA$, 131 $\rm\AA$ and 171 $\rm\AA$ marked by the vertical red arrows and the later four blue vertical dashed lines. In other words, several EUV jets also appeared in the vicinity of the filament south-western footpoint during period of the flux emergence. The characteristic emission temperature of the AIA-304 $\rm\AA$ band and the one of AIA-171 $\rm\AA$ band are around $10^{4.7}$ K and $10^{5.8}$ K. However, the emission contribution of AIA-131 $\rm\AA$ band are from two ions, which is Fe VIII around at $10^{5.6}$ K and Fe XX around at $10^{7.0}$ K. In non-flaring conditions, emission from the very hot Fe XX ion is usually negligible \citep{mar11}. Moreover, the characteristic emission temperatures of other EUV bands are higher than $10^{6.0}$ K. This means that these EUV jets are more sensitive in lower emission temperature. It should be noted that all the intensities of $SDO$/AIA bands did not enhance during the jets recorded by the NVST at 05:09 UT and 05:59 UT. These two jets are indicated by two red vertical dash lines in the Fig.\ref{fig5} (b). Moreover, these two jets can be considered as H$\alpha$ jets which only see in the H$\alpha$ band. On the other hand, the new emerging flux had been occurred before the onset of H$\alpha$ jets (see Fig.\ref{fig5} (a) and (b)), which begin to emerge at around 04:58 UT. Therefore, it is reasonable suspected that these new emerging flux result in the occurrences of these H$\alpha$ jets or EUV jets. The new emerging flux could reconnect with the pre-existing magnetic field, which maybe cause the occurrences of these jets. However, some H$\alpha$ jets could not be detected by the $SDO$/AIA instrument, which may be due to the low temperature of the H$\alpha$ jets and could not enhance the intensities of $SDO$/AIA band lines. Nevertheless, these cool H$\alpha$ jets also could lift the cool plasma into the filament height, and supply the material for the filament formation.

\begin{figure}
\includegraphics[width=\columnwidth]{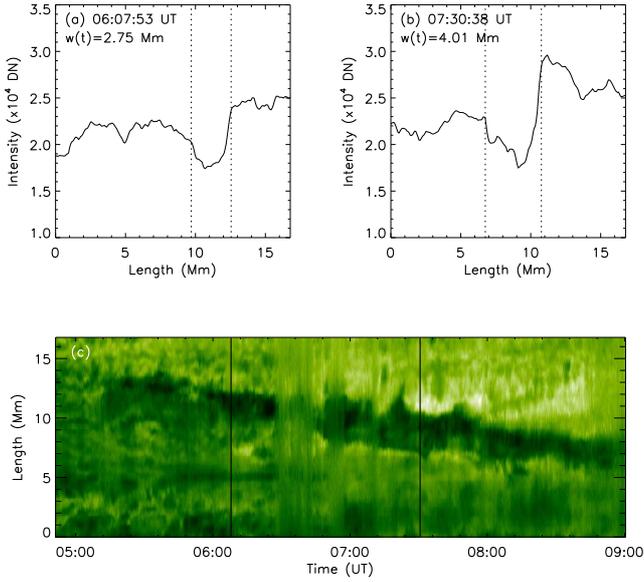}
\caption{Time-distance diagram along the blue dashed lines in the panel (b) of Fig.\ref{fig4}. (a) Variation of the H$\alpha$ intensity along the slice path at 06:07:53 UT. (b) Variation of the H$\alpha$ intensity along the slice path at 07:30:38 UT. The vertical dotted lines denote the boundaries of jet plasma. (c) The time-distance diagram during the period from 04:48 UT to 09:00 UT. The two vertical black lines indicate the time-distance diagram positions of panels (a) and (b), respectively. \label{fig6}}
\end{figure}

In order to estimate the mass of the cool plasma driven by the jets, we assume the cross-section of the jets is circular. The quantity of injection mass can be calculated by the following equation:
\begin{equation}\label{eq5}
    M_j=\int m_H n_H \frac{w^2(t)\pi}{4}v(t)dt,
\end{equation}
where $m_H$ is the mass of the hydrogen atom, $n_H$ is the total hydrogen number density, $w(t)$ is the width of the cross-section of the jet, $v(t)$ is the speed of the cool plasma. Considering that the cool plasma come from the lower atmosphere, it is reasonable to assume that the density of the cool plasma driven by the jets is similar to that of the chromosphere. Thus, the total hydrogen number density $n_H$ is about 3.26 $\times$ $10^{10}$ $cm^{-3}$. The mass of the hydrogen atom $m_H$ is about 1.67 $\times$ $10^{-24}$ g. From the above diagnosis, the speed of the cool plasma ($v(t)$) was estimated as about 20-30 km/s. In order to derive the width cross-section of the jets ($w(t)$), we make a time-distance diagram along the blue dashed line in the panel (b) of Fig.\ref{fig4}. To allow for uncertainty in the definition of the boundary between the jet and the filament, the blue dashed line was placed nearby the footpoints of the jets. This avoids the effect of the main filament as much as possible. According to the animation of Fig.\ref{fig1}, the cool material went across this place continuously during the period about from 05:10 UT to 08:54 UT and then the filament appeared. On the other hand, the jet was wider than the filament, so the jet's width would embody the filament. Therefore, the uncertainty in the definition of the boundary between the jet and the filament had only little effect on the estimate of the cross-sectional width of the jets. Fig.\ref{fig6} exhibits the time-distance diagram derived by a series of H$\alpha$ images observed by the NVST during the period from 04:48 UT to 09:00 UT.  Panels (a)\&(b) show the variations of intensities along the slice path, respectively. Using the different intensities along the slice and the maximum of the variation rate nearby the jets' position, we can approximately identify the two boundaries of the jet plasma to derive the width of the cross-section. For the complex profiles of the slice's intensity, it need manually identify the boundary. We obtain the $w(t)$ is around 2.75 Mm at 06:07:53 UT, while the $w(t)$ is around 4.01 Mm at 07:30:38 UT. Therefore, based on the time-distance diagram, the $w(t)$ at different moments can be derived by using the same method. According to the Eq.\ref{eq5} and above parameters of the jets, the total mass ($M_j$) of the cool plasma carried by jets is estimated to be in the range (9.3-14.0) $\times$ $10^{13}$ g. On the other hand, the filament is assumed as a circular slab. The mass of the filament $M_f$ also can be estimated by following equation:
\begin{equation}\label{eq6}
    M_f=n_Hm_H\frac{w_f^2\pi}{4}L_f,
\end{equation}
where $n_H$ is the total hydrogen density of the filament, $w_f$ is the width of the filament, $L_f$ is the length of the filament. From Fig.\ref{fig1}, we obtain the filament width ($w_f$) is about 2.0 Mm and the filament length ($L_f$) is about 23.7 Mm. The total hydrogen density $n_H$ in the filament measured ranges about from $3\times 10^{10}$ $cm^{-3}$ \citep{ste97} to $3\times 10^{11}$ $cm^{-3}$ \citep{hir86}. Based on the values, the total filament mass ($M_f$) is estimated to be in the range (0.37-3.7) $\times$ $10^{13}$ g. Comparing the $M_j$ with $M_f$, the maximum of the $M_f$ is just about 40$\%$ of the minimum of the $M_j$. Although not all the cool plasma ejected by jets would become the filament material and some of the cool plasma would fall down to the chromosphere, a conservative estimate of the mass carried by jets could supply sufficient material for the filament at least.

\subsection{The magnetic properties of the newly emerging flux}
\begin{figure}
\includegraphics[width=\columnwidth]{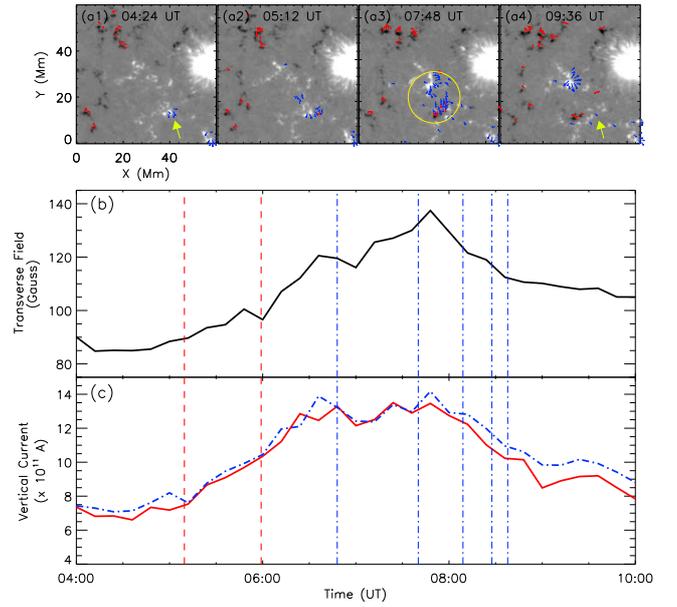}
\caption{(a1)-(a4) Vector magnetograms from $SDO$/HMI. The blue arrows denotes the transverse magnetic field sited in the vertical magnetic fields with positive polarity, while the red ones denotes the transverse magnetic field sited in the vertical magnetic fields with negative polarity. (b) The time variation of mean transverse field in the yellow circle. (c) The variations of the positive and negative  vertical current. The red solid line denotes the time profile of the positive current, while the blue dashed-dotted line denotes the time profile of the negative current. The two red vertical dashed lines indicate the onset of two H$\alpha$ jets, while the blue dashed-dotted lines indicate the moments of EVU jets.\label{fig7}}
\end{figure}

From the Fig.\ref{fig5} (a) and the analyze in the section.\ref{sec3.2}, we find some newly emerging flux presented in vicinity of the filament south-western footpoint during the formation of the filament. Previous investigations have shown that many jets have a close relationship with newly magnetic flux emergence \citep{shi98,rao16,hon17}. Thus, these newly emerging flux may play important role in the jets and the magnetic structure of the filament. In other words, these newly emerging flux may be a source of supplying mass and energy for the filament. In order to investigate the properties of newly emerging flux associated with the formation of the filament, we calculate the transverse field and vertical current in the photosphere in the newly flux emergence area. Figs.\ref{fig7} (a1)-(a4) show vector magnetograms at different moments observed by $SDO$/HMI. The background denotes the vertical magnetic field, while the blue and the red arrows denote the transverse magnetic field with the positive and negative vertical magnetic field, respectively. From the first three magnetograms, it is noted that some newly fluxes emerged in the yellow circle in the panel (a3), in which the vertical and transverse magnetic fields enhanced. And then, the newly emerging flux canceled with the surrounding magnetic field marked by the yellow arrows in the panels (a1)\&(a4). Panel (b) of the Fig.\ref{fig7} shows the time variation of the mean transverse field strength in yellow circle during the period from 04:00 UT to 10:00 UT. The mean transverse field strength rapidly increased from 85 G to 135 G during the period from about 05:00 UT to 07:48 UT. The mean rate of the increasing is about 18.2 G/h. After that, it decreased in two phases. Firstly, it decayed rapidly from 135 G to 112 G during the period from 07:48 UT to 08:36 UT, and the mean rate of decaying is 28.7 G/h. Secondly, it decayed slowly from 112 G to 105 G during the period from 08:36 UT to 10:00 UT, and the mean rate of decaying is about 5 G/h. With the describing in the section \ref{sec:obser methods}, we can obtain the vertical current in the yellow circle. The panel (c) exhibits the time profiles of the positive and the negative vertical current in yellow circle, in which the positive vertical current is marked by the red solid line and the negative vertical current is marked by the blue dashed-dotted line. Both of the positive and negative vertical currents display a similar temporal evolution and have relatively equivalent quantity. They had a quickly increase from 7 $\times$ $10^{11}$ A to 14 $\times$ $10^{11}$ A during the period from 05:00 UT to 06:36 UT. The mean rate of the increase is 4.7 $\times$ $10^{11}$ A/h. During the period from 06:36 UT to 07:36 UT, both of them kept at a relatively high value and had a slightly turbulent temporal variations. After that, they decreased gradually to about 8 $\times$ $10^{11}$ A. Obviously, two non-potential proxies of the magnetic field (the transverse field and current) would also enhanced at the former phase of the flux emergence. Comparing the EUV jets, the two H$\alpha$ jets occurred with the relatively less transverse field and vertical current. This may be due to the less energies accumulated at the former phase of the flux emergence. It is too weak to cause the EUV jets' happening.

\begin{figure}
\includegraphics[width=\columnwidth]{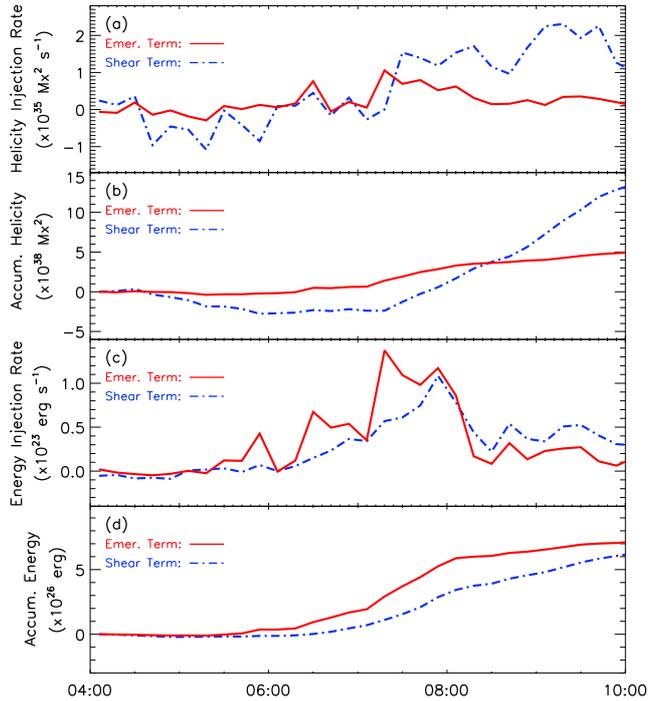}
\caption{Time variation curves of different physical parameters across the yellow circle in the panel (a3) of Fig.\ref{fig7}. (a) Helicity injection rate. (b) Accumulative helicity. (c) Energy injection rate. (d) Accumulative energy. The red solid lines denote the quantities from the emergence term, while the blue ones denote the quantities from the shear term. \label{fig8}}
\end{figure}

With the method described in section \ref{sec:obser methods}, we also calculate the magnetic helicity injection rate and magnetic energy injection rate (Poynting vector) from the solar interior to atmosphere in the yellow circle in the panel (c) of Fig.\ref{fig7}. We set the accumulative helicity and accumulative energy to zero at the beginning of investigation. Fig.\ref{fig8} shows the time variations of helicity injection rate, accumulative helicity, energy injection rate, accumulative energy in different panels, respectively. The blue dotted-dashed lines denote the quantities from the shear term, while the red solid lines denotes the quantities from the emergence term. From the panel (a), the helicity injection rate from the emergence term ($\rm \dot{H}_e$) is about zero before the flux emergence and remains positive injection during the flux emergence, while the helicity injection rate from the shear term ($\rm \dot{H}_s$) is negative at the early phase and increases to positive after around 07:00 UT. In general, the value of $\rm \dot{H}_s$ is larger than that of $\rm \dot{H}_e$. Most of the magnetic helicity injection is contributed by the shear term, which also is demonstrated by panel (b). The accumulated helicity from the shear term ($\rm H_s$) is up to 13 $\times$ $10^{38}$ $\rm Mx^2$ at 10:00 UT, while the one from the emergence term ($\rm H_e$) is only 5 $\times$ $10^{38}$ $\rm Mx^2$. Panel (c) exhibits time profiles of the energy injection rates from emergence term ($\rm \dot{E}_e$) and shear term ($\rm \dot{E}_s$). Two term energy rates have positive, which manifests that the magnetic energy are injected from subsurface to atmosphere. Both of energy rates abruptly enhanced during the flux emergence and had a sharply decrease after the flux emergence. The value of $\rm \dot{E}_e$ is sightly larger than that of $\rm \dot{E}_s$, which means that the magnetic energy injection are contributed by both terms. Panel (d) shows the time variations of the accumulative magnetic energy from both terms. The magnetic energy from two terms increased quickly from around 05:30 UT to 07:48 UT and then began slowly increasing. The total magnetic energy injection could accumulated up to about $10^{27}$ erg. On other hand, combining the Fig.\ref{fig7}, the two H$\alpha$ jets took place at the moments with relative lower accumulated energy while the EUV jets took place at the moments with relative higher accumulated energy. The newly emerging flux carried the magnetic helicity and magnetic energy into the atmosphere nearby the western footpoint of the filament, which would provide the energy to carry the mass for the filament and the nonpotential field (helicity) to rearrange the magnetic structure for the filament.
\section{Conclusion and discussion}\label{sec:conclusion}
In this paper, we have presented and studied the formation process of a filament in active region NOAA 11903 in detail and provide a new view of understanding the material supply of the solar filament. The complete formation process of the filament was captured by the NVST and SDO observations, which gradually formed from nonexistence of filament at the beginning of the investigation. We mainly investigate the material injection of the filament and analyzed the newly emerging flux related to the material injection nearby the one footpoint of the filament. The main results are as follows.

1. The material of the filament originates from the chromosphere. Thought the influence of the instabilities or solar activities, the plasma in the chromosphere could be lifted into the upper atmosphere and some of them become the filament material.

2. The jets occur in the vicinity of the south-western footpoint of the filament and the newly emerging flux can be found in the same place. Thus, the new flux emergence is responsible for these jets, which were caused by the magnetic reconnection between the pre-existing and new emergence magnetic field. The total mass carried by the jets is about (9.3-14.0) $\times$ $10^{13}$ g, while the mass of the filament is estimated to be in the range (0.37-3.7) $\times$ $10^{13}$ g. Jet is the sufficient way to upload the cool plasma and supply the material for the filament.

3. Although some H$\alpha$ jets (cool jets) could not brighten the intensity of the SDO/AIA bands due to magnetic reconnection with the low temperature or in the low height, but these cool jets also could lift the cool plasma (T $\sim$ $10^4$ K) to filament height and supply the material for the filament. The velocities of the projection ejected cool plasma range about from 20-30 km/s by the jets. Cool jet is also an important way for supplying the material for the filament.

4.The newly emerging flux carried the energy and helicity into atmosphere nearby the western footpoint of the filament. It may play an important role in transporting the material for filament and building the magnetic structure for the filament.

5. The H$\alpha$ jets occurred at the beginning of the magnetic emergence, while the EUV jets occurred at the latter of the magnetic emergence with larger accumulated energy.

Furthermore, the twisted structures can be identified in the one filamentary structure during the material injection events. This filamentary structure with high twisted structures eventually evolved as the filament, while the another filamentary structure could not maintain stable as the ejected plasma falls down the chromosphere. \cite{wan18} also suggested that the ejected plasma is more earlier to be captured by the high twisted structures magnetic field and more stable to remain in the filament. In addiction, previous studies also showed that high twisted structures could be found in the filament during the activity or eruption of solar filaments \citep{che14b,yang14,bi15,xue16,jia16}. Therefore, the twisted structures (or magnetic dips) play an important role in holding the mass for the filament.

It is widely accepted that the material of the filaments or prominences should originate from the low atmosphere (chromosphere) \citep{pik71,zir94,liu05,yan16b,wan18}. In this paper, we also confirm that the mass of the filament was extracted from the low atmosphere instead of the corona. However, the way how the plasma in the low atmosphere lifts into the filament or filament channel puzzles solar researchers for a long time. Many authors proposed different models for trying to understand the physical mechanisms in the transport process of the filament material \citep{mac10,kar15}. The evaporation-condensation models and injection models are more popular among of them. Many numerical simulations manifest that the mass could be condensed near the apex when heat the plasma near the footpoint of the loop \citep{ant99,kar03,kar05,kar08,xia11,lun12}, which are related to thermal instability \citep{par53,fie65}. However, it is hardly to explain the source of the heating. Furthermore, the only few observation evidences could be captured to supported to the this models \citep{ber12,liu12}. On the other hand, many observations are more likely to support the injection models which rely on magnetic reconnection in the low atmosphere to propel cool plasma to typical filament height \citep{wan99,liu05,zou17,wan18}. In this paper, we supplied a observational evidence that the filament material can be supplied by a series of jets related to the new emerging magnetic field, which again verify the injection model. The jets directly lifted the cool plasma in the low atmosphere to the filament height, while some of the lifted cool plasma could remain in the filament height and became the mass of the filament. Therefore, jet is the sufficient way to upload the cool plasma in the chromosphere into the filament. We also find that H$\alpha$ or cool jets are also enough strong to carry the material for the filament formation, not just EUV jets. In our opinion, it may open a new window for understanding the material supply of the filament. Because there are many small explosive events in the lower atmosphere. These small explosive events also would carry the cool plasma to filament height, not just big explosive events. This means that some filament material could come from the accumulation of the cool plasma carried by these small explosive events.

Using the H$\alpha$ images observed by NVST, we estimate the the mass ($M_j$) of cool plasma ($T\sim10^4$ K) carried by jets and the mass ($M_f$) of the filament and find that the jets can supply the sufficient mass for the filament. In the investigation of \cite{cha03}, the authors used EUV data to estimate the mass of EVU-emitting plasma ($T\sim$ 2-3 $\times$ $10^5$ K) in the jets and eruptions and found the total mass in the EUV jets and eruption was comparable with the mass of the filament. And the authors also indicated the existence of such cool material could be lifted in these jets and eruptions. In our study, some H$\alpha$ jets also could lifted the cool plasma to filament height. This means not all the jets carrying the material for the filament could enhance the brightening of EVU bands. In the other hand, hot plasma in the jets often have higher speed and momentum \citep{she17}. These hot plasma would be hardly to remain in the filament. Therefore, we consider that the cool plasma ($T\sim10^4$ K) directly injected into the filament from the chromosphere is dominant component of supplying the material for solar filaments.

\begin{figure}
\includegraphics[width=\columnwidth]{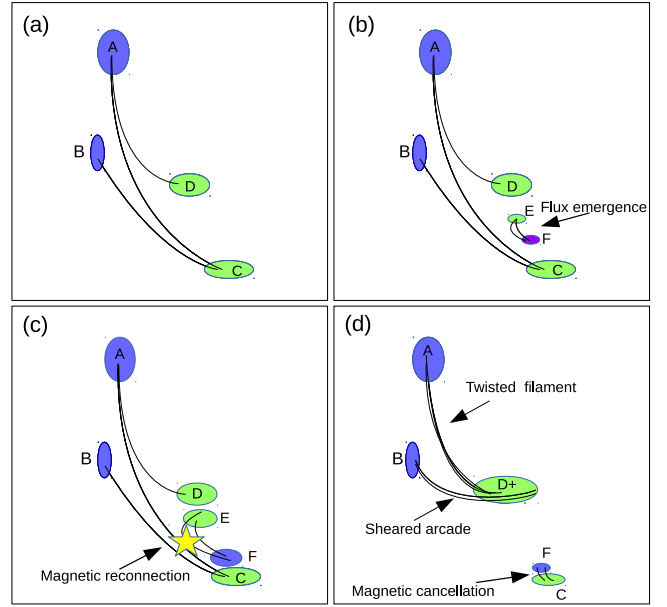}
\caption{Cartoon showing the process of magnetic reconnection between newly emerging flux and pre-exsited magnetic field. The blue filled ellipses denote the negative magnetic polarities, while the green filled ellipses denote the positive polarities.\label{fig9}}
\end{figure}
In this study, we find that some newly emerging flux occurred in the vicinity of the south-western footpoint of the filament during the jets. Based on many MHD simulations of jets, the jets can be caused by the magnetic reconnection between an emerging magnetic flux and the pre-existing magnetic field \citep{yok96,mor08,mor13,cheu14}. Therefore, we considered that this jets are caused by the magnetic reconnection between the new emerging flux and the pre-existing magnetic field. According to the observational characteristics, we drawn a cartoon to demonstrate this process (See Fig.\ref{fig9}). The blue filled ellipses denote the negative magnetic polarities, while the green filled ellipses denote the positive polarities. Panel (a) of Fig.\ref{fig9} shows the original magnetic configuration at the beginning. Some new flux (EF) emerged in the vicinity of the positive polarity C (Fig.\ref{fig9} (b)). As this newly emerging flux rose, the new magnetic fields (EF) and the pre-existing magnetic fields (AC \& BC) reconnected with each other (Fig.\ref{fig9} (c)). The positive polarities D and E merged together, which became the new positive polarity marked by symbol D+. And then, two magnetic fields (AD+ and BD+) were formed, which correspond to the two filamentary structures what we observed. In the meanwhile, magnetic cancellation was caused by the submergence of post-reconnection field (FC). It is worth noting that it can not be deduced how the twist in the filament originated using our observations. There are three possible ways. One is that the twist had existed in the pre-existing magnetic field before the flux emergence. The second possible way is that the twist had existed in the newly emerging flux and transported to the filament during the magnetic reconnection. Another possibility, is that the twist formed during the magnetic reconnection. It would need more observations and more study to address this issue.

It is easy to understand that different non-potential parameters (e.g., transverse field, vertical electric current) are increasing during the the magnetic flux emergence. After all, these parameters are the proxies of the non-potential magnetic field. On the other hand, H$\alpha$ jets occurred in the week emerging flux, and the EUV jets occurred in the strong emerging flux. Furthermore, the two H$\alpha$ jets occurred at the relatively less transverse magnetic field and low vertical current. This may indicate that the magnetic reconnection between week emerging flux with less energy and the pre-existing magnetic field caused the H$\alpha$ jets. The magnetic reconnection between strong emerging flux with high energy and the pre-existing magnetic field caused the EUV jets.

In addition, we derive the accumulated energy carried by the newly emerging flux is about $10^{27}$ erg. We can simply estimate the energy required for lifting the jets mass from chromosphere to filament height as following equation: $\rm E_j=M_jg_\odot \Delta h $, where the $g_\odot$ is the Sun acceleration of gravity and the $\Delta h$ is the lifting height of the mass. $g_\odot$ is 2.74 $\times$ $10^2$ m/s on the solar surface. And we assume $\Delta h$ equal to the typical height of active region filament from chromosphere about 2 Mm. Thus, the energy required for lifting the jets mass from chromosphere to filament height ($\rm E_j$) ranges about (0.51-0.76) $\times$ $10^{27}$ erg. Although additional heat and light energy could also be consumed during the jets, the newly emerging flux could provide the energy for lifting the mass for the filament.

On the acceleration mechanisms of cool plasma during the jets, previous studies found that magnetic reconnection in the lower atmosphere can generation slow mode wave \citep{tak01} and as a slow-mode wave propagates upward, its amplitude grows due to the density contrast, eventually steepening into a shock. The interaction of the shock and the transition region between the chromosphere and the corona can launch the transition region upward, which is observed as a cool jet \citep{shi82,pon04,heg09}. Furthermore, using a two-dimensional magnetohydrodynamic (MHD) simulation, \cite{Tak13} also found that slow-mode waves produced in magnetic reconnection would play key roles in the acceleration mechanisms of chromospheric jets and the chromospheric plasma could be accelerated due to the combination of the Lorentz force and the whip-like motion of the magnetic field when magnetic reconnection takes place in the upper chromosphere. The authors suggested three types of acceleration mechanisms of cool plasma: shock acceleration type, shock and whip-like acceleration type, and whip-like acceleration type \citep{yok96} according to the hight of the magnetic reconnection site and found that the magnetic energy released by magnetic reconnection is efficiently converted into the kinetic energy of jets in these process. The further details and realism would need to be studied using three-dimensional simulations.

Thanks to the high spacial and temporal resolution observations from the NVST, some low temperature H$\alpha$ jets were found. These H$\alpha$ jets could not brighten in the $SDO$/AIA EUV bands, but they also could eject the cool plasma to the filament height and supply the material for the filament. This attractive phenomenon impels us to suspicions that the mass of all filaments may be related to the different types of jets. Some jets could be strong enough to be observed by recent telescopes, while other cool jets (min-jets) could not be observed due to the low temperature or low intensity. This extraordinary idea needs more observation evidences to confirm, which maybe need more higher spatial and temporal resolution and different layer observations provided by much bigger aperture telescopes in the future.
\section*{Acknowledgements}
The authors thank the referee for constructive comments and suggestions  that  have improved the  quality of the manuscript. SDO is a mission of NASA's Living With a Star Program. The authors are indebted to the SDO, NVST teams for providing the data. This work is supported by the National Science Foundation of China (NSFC) under grant numbers 11873087, 11503080, 11603071, 11633008, 11527804, the Yunnan Talent Science Foundation of China, (2018FA001) the Youth Innovation Promotion Association CAS under number 2011056, the CAS ``Light of West China" Program under number Y9XB018001, the grant associated with project of the Group for Innovation of Yunnan province.

\end{document}